\documentclass[sigconf,natbib=true]{acmart}

\usepackage{graphicx}
\usepackage{subfiles}

\usepackage{todonotes}
\usepackage{booktabs}
\usepackage{multicol}
\usepackage{multirow}
\PassOptionsToPackage{hyphens}{url}\usepackage{hyperref}

\usepackage{dsfont}
\usepackage{amsmath}
\usepackage{bbm}

\usepackage{pifont}
\newcommand{\cmark}{\ding{51}}%
\newcommand{\xmark}{\ding{55}}%

\usepackage{caption}
\usepackage{subcaption}

\usepackage{xcolor}

\AtBeginDocument{%
  \providecommand\BibTeX{{%
    \normalfont B\kern-0.5em{\scshape i\kern-0.25em b}\kern-0.8em\TeX}}}

\copyrightyear{2022}
\acmYear{2022}
\setcopyright{acmcopyright}\acmConference[SIGIR '22]{Proceedings of the 45th International ACM SIGIR Conference on Research and Development in Information Retrieval}{July 11--15, 2022}{Madrid, Spain}
\acmBooktitle{Proceedings of the 45th International ACM SIGIR Conference on Research and Development in Information Retrieval (SIGIR '22), July 11--15, 2022, Madrid, Spain}
\acmPrice{15.00}
\acmDOI{10.1145/3477495.3531886}
\acmISBN{978-1-4503-8732-3/22/07}

\begin{document}

\title[C3: Continued Pretraining with Contrastive Weak Supervision for Cross Language Ad-Hoc Retrieval]{C3: Continued Pretraining with Contrastive Weak Supervision \\ for Cross Language Ad-Hoc Retrieval}

\author{Eugene Yang}
\affiliation{%
  \institution{HLTCOE, \\ Johns Hopkins University, USA}
  \country{}}
\email{eyang35@jhu.edu}

\author{Suraj Nair}
\affiliation{
    \institution{University of Maryland, USA}
    \country{}
}
\email{srnair@umd.edu}

\author{Ramraj Chandradevan}
\affiliation{\institution{Emory University, USA}\country{}}
\email{rchand31@emory.edu}

\author{Rebecca Iglesias-Flores}
\affiliation{
    \institution{University of Pennsylvania, USA}
    \country{}
}
\email{irebecca@seas.upenn.edu}

\author{Douglas W. Oard}
\affiliation{
    \institution{University of Maryland, USA}
    \country{}
}
\email{oard@umd.edu}

\renewcommand{\shortauthors}{Yang, et al.}

\begin{abstract}
Pretrained language models have improved effectiveness on numerous tasks, including ad-hoc retrieval. Recent work has shown that continuing to pretrain a language model with auxiliary objectives before fine-tuning on the retrieval task can further improve retrieval effectiveness. 
Unlike monolingual retrieval, designing an appropriate auxiliary task for cross-language mappings is challenging. 
To address this challenge, we use comparable Wikipedia articles in different languages to further pretrain off-the-shelf multilingual pretrained models before fine-tuning on the retrieval task. We show that our approach yields improvements in retrieval effectiveness. 
\end{abstract}



\maketitle

\section{Introduction}

\begin{figure}
    \centering
    \includegraphics[width=\linewidth]{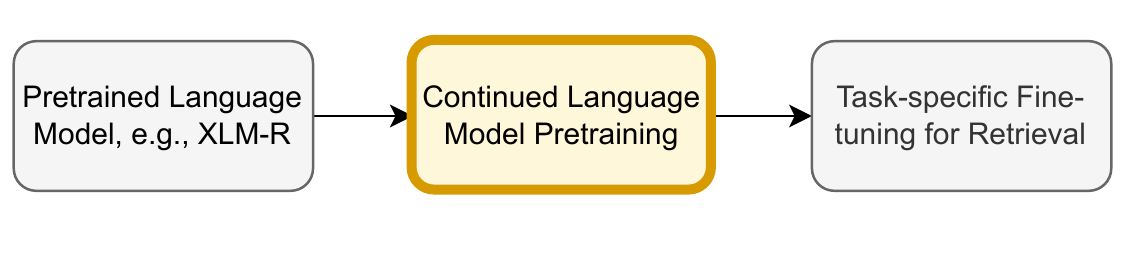}
    \vspace{-2.5em}
    \caption{Pipeline for training a dense retrieval model. We propose an additional pretraining phase targeting CLIR. }\label{fig:pretrain-pipeline}
\end{figure}

Dense retrieval models, such as ColBERT~\cite{khattab2020colbert}, ANCE~\cite{xiong2020approximate}, and DPR~\cite{karpukhin2020dense}, have been adapted to cross language ad-hoc retrieval (CLIR) where queries and documents are in different languages by replacing monolingual embedding with a multilingual embeddings (e.g., mBERT~\cite{devlin-etal-2019-bert} and XLM-R~\cite{conneau2020unsupervised}). 
These dense retrieval models learn to encode queries and documents separately into fixed-length dense representations by fine-tuning a pretrained model (e.g, BERT~\cite{devlin-etal-2019-bert}) with a retrieval objective using a large number of query-document pairs such as the ones from MS MARCO~\cite{bajaj2018ms} or Natural Questions~\cite{kwiatkowski-etal-2019-natural}.
Recent work showed that these models are effective for CLIR when trained with monolingual query-document pairs, enabling zero-shot transfer~\cite{colbertx, macavaney2020newdog, shi2019cross}. Alternatively, training the model with translated MS MARCO (translate-train) is more effective but also much more expensive~\cite{colbertx,shi2021cross}.

However, most pretrained language models do not explicitly ensure the representations of a pair of related texts are similar~\cite{gao2021condenser}. This calls for a task-specific fine-tuning process to retrofit the representation produced by the pretrained model to be closer between related or relevant text. Such processes can be complex and computationally expensive, such as RocketQA~\cite{qu2020rocketqa}, and, thus, efficient multi-stage training that introduces a ``continued pretraining'' to the pipeline was proposed for monolingual retrieval~\cite{gao2021unsupervised,izacard2021unsupervised} before running a task-specific fine-tuning with retrieval objectives (illustrated in Figure~\ref{fig:pretrain-pipeline}).

By construction, the representations for the text conveying similar information in different languages are not necessarily similar, since multilingual pretrained models such as mBERT and XLM-R do not introduce parallel text during pretraining. 
In other settings, incorporating alignment information into the retrieval model has been shown to be useful for CLIR~\cite{huang2021mixed,darwish2003psq}. We hypothesize that explicitly promoting token-level similarity during the pretraining phase will enhance the effectiveness of CLIR models.

To address the aforementioned issues, we propose C3, a continued pretraining approach leveraging weak supervision with document-aligned comparable corpora to encourage representations of the text with similar meaning in different languages to be more similar using contrastive learning. This continued pretraining phase modifies an off-the-shelf pretrained model before it is fine-tuned to the actual retrieval objective, as illustrated in Figure~\ref{fig:pretrain-pipeline}.
Specifically, we model the similarity between a pair of texts using contrastive learning with token-level embeddings to encourage the model to embed token-level similarity and alignment information. We use Wikipedia articles in the language pair of interest, linking them based on cross-language links present in Wikipedia. We test this using high-resource languages for which strong evaluation resources are available. but this weakly supervised approach could also be applied to lower resource languages in which alternative approaches that rely on parallel text might prove impractical. 

To summarize, our contributions are two-fold. First, we propose to continually pretrain the model for CLIR dense retrieval to promote similar representations between texts with similar meanings in different languages, using contrastive learning. To our best knowledge, this is the first work that applies contrastive learning to CLIR pretraining in this way. Secondly, we do this in a way that relies only on weakly supervised document-scale links in Wikipedia.

\section{Background and Related Work}

Since the introduction of pretrained transformer-based language models, neural retrieval models have been taking advantage of these models for more effective query-document matching. Early work in monolingual retrieval involved building cross-encoder models ~\cite{Dai2019-sl, macavaney2019cedr, nogueira2019multistage} that leveraged the full interaction between the queries and documents to produce the relevance scores. Subsequently, similar models ~\cite{Shi2019-on, zhao2019weakly, Jiang2020-dt, Yu2020-dl, huang2021mixed} were adapted to the CLIR setting. While effective, such models can only rerank documents since they process both queries and documents during inference and thus yield a longer running time compared to traditional sparse retrieval techniques such as BM25~\cite{Robertson1994OkapiAT}. DPR-style dense retrieval models overcome this limitation by scoring the documents based on the similarity of their representations, which allows the language models to encode documents beforehand. 

However, the representations produced by off-the-shelf language models are undertrained~\cite{gao2021unsupervised} and thus directly scoring documents with such representations yields suboptimal retrieval results~\cite{litschko2021cross}. Additional task-specific fine-tuning with relevance labels produces much better representations on either the sequence level, such as ANCE~\cite{xiong2020approximate} and DPR~\cite{karpukhin2020dense}, or the token level, such as ColBERT~\cite{khattab2020colbert}. Especially for sequence level representations, often summarized in the \texttt{CLS} token, contrastive learning~\cite{oord2018infonce,chen2020simclr} that trains a model with one positive and multiple negative examples for each query has been shown to be one of the most effective training techniques for dense retrievers~\cite{qu2020rocketqa, xiong2020approximate, izacard2021towards}. In-batch negative sampling further reduces the memory requirement by treating positive examples for other queries as negative~\cite{gao2021unsupervised,karpukhin2020dense}. A similar objective was utilized to pretrain the cross-encoder retrieval model for the CLIR task ~\cite{yu2021cross}.

Continuing pretraining the off-the-shelf language model has been investigated in mono-lingual retrival~\cite{chang2020pre, fan2021pre, gao2021unsupervised}. 
Specifically, coCondenser~\cite{gao2021unsupervised} continued pretraining of the language model with a passage-containing classification task (i.e., determining if a pair of passages belong to the same document) through contrastive learning on the representation of the passages for monolingual IR before fine-tuning it as a DPR model. This weakly supervised pretraining teaches the model to bring the representation of passages extracted from the same documents closer, which benefits the downstream dense retrieval task by assuming passages from the same document convey a similar meaning. coCondenser also trains with a masked language modeling task on the Condenser head~\cite{gao2021condenser} that adds two additional layers at the end of the network with the embeddings of \texttt{CLS} from the last layer and the rest from the middle layer. This Condenser head is removed after pretraining, but it has been shown to adjust the encoder effectively. 

Recently, dense retrieval models have been adapted to CLIR by replacing the encoder with a multilingual pretrained model, such as mBERT, XLM or XLM-R~\cite{Bonab2020-xk, colbertx, litschko2021cross}. To utilize existing monolingual collections with a large number of relevance labels such as MS MARCO~\cite{bajaj2018ms}, dense retrievers with multilingual embeddings can be trained on such corpora with zero-shot transfer to CLIR by leveraging the multilinguality of the encoder~\cite{colbertx, macavaney2020newdog}. Alternatively, with the help of translation models, one can translate the monolingual training collection into the language pair of interest and train the retriever on it (a ``translate-train'' approach)~\cite{colbertx,Shi2019-on}. This training scheme encourages the model to bring the representations of related queries and documents closer across languages. However, training effective translation models can be resource-intensive.

Besides the challenges in obtaining the translation, teaching the models two complex tasks jointly can also be tricky. Learning from coCondenser, a two-stage process with a continued language model pretraining followed by task-specific fine-tuning can help the model acquire knowledge incrementally. The following section introduces a pretraining approach that encourages the model to bring the representations of passages in different languages with similar meanings closer before fine-tuning with retrieval objectives. 
\begin{table*}[t]
\renewcommand{\b}[1]{\textbf{#1}}
\renewcommand{\d}{$\dagger$}

\caption{Reranking effectivness of ColBERT and DPR models with and without our C3 pretraining. The top shows XLM-RoBERTa-base models; the bottom shows XLM-algin-base models. Symbols indicate statistically significant differences at $p<0.05$ by a two-tailed paired $t$-test with Bonferroni correction for 6 tests, either with and without C3 (*) or between C3 and original BM25 results (\d). $\Delta$ shows the mean relative improvement from C3 across the 6 collections.} \label{tab:main-results}
\centering
\resizebox{\textwidth}{!}{%
\begin{tabular}{lc|cc|c|ccc|c||cc|c|ccc|c}
\toprule
      &            & \multicolumn{7}{c||}{nDCG@100}  &  \multicolumn{7}{c}{nDCG@10} \\
\midrule
Retrieval & With      & \multicolumn{2}{c|}{HC4} & \multicolumn{1}{c|}{NTCIR} & \multicolumn{3}{c|}{CLEF} &&
                        \multicolumn{2}{c|}{HC4} & \multicolumn{1}{c|}{NTCIR} & \multicolumn{3}{c|}{CLEF} &\\
Model     & C3        & Chinese & Persian & Chinese & Persian & German & French & $\Delta$&
                        Chinese & Persian & Chinese & Persian & German & French & $\Delta$\\
\midrule
\multicolumn{2}{c|}{QT + BM25} 
           &   0.362 &   0.354 &  0.264 &    0.336 &   0.419 &   0.563 &&    0.258 &   0.251 &   0.229 &    0.407 &    0.379 &   0.505 &\\
\midrule
\multicolumn{16}{l}{XLM-RoBERTa (base)}\\
\midrule
\multirow{2}{*}{ColBERT}
& \xmark &    0.352 &   0.385 &   0.249 &      0.283 &     0.510 &   0.590 &&    0.248 &   0.277 &   0.223 &    0.325 &     0.513 &   0.514 &\\
& \cmark &\b{*0.444}&\b{0.391}&\b{0.278}&\b{\d*0.286}&\b{\d0.521}&   0.574 &+8\%&\b{*0.345}&   0.274 &\b{0.255}&\b{ 0.337}&\b{\d0.535}&   0.482 &+11\%\\
\midrule
\multirow{2}{*}{DPR}
& \xmark &    0.330 &   0.319 &   0.218 &     0.259 &     0.467 &   0.531 &&    0.223 &   0.220 &   0.184 &     0.299 &      0.449 &   0.449 &\\
& \cmark &\b{*0.395}&\b{0.341}&\b{0.255}&\b{\d0.266}&\b{\d0.503}&\b{0.562}&+10\%&\b{*0.287}&\b{0.226}&\b{0.231}&\b{\d0.302}&\b{\d*0.523}&\b{0.491}&+15\%\\
\midrule
\multicolumn{16}{l}{XLM-align (base)}\\
\midrule
\multirow{2}{*}{ColBERT}
& \xmark &      0.425 &   0.399 &     0.303 &    0.252 &     0.523 &   0.579 &&      0.332 &   0.294 &     0.283 &    0.285 &     0.532 &   0.478 &\\
& \cmark &\b{\d*0.483}&\b{0.400}&\b{\d0.330}&\b{ 0.275}&\b{\d0.528}&\b{0.588}&+4\%&\b{\d*0.408}&   0.280 &\b{\d0.316}&\b{ 0.321}&\b{\d0.536}&\b{0.499}&+6\%\\
\midrule
\multirow{2}{*}{DPR}
& \xmark &    0.385 &   0.366 &   0.260 &     0.235 &     0.480 &   0.581 &&    0.300 &   0.256 &   0.239 &    0.265 &      0.482 &   0.503 &\\
& \cmark &\b{ 0.421}&\b{0.403}&\b{0.286}&\b{\d0.244}&\b{\d0.503}&\b{0.586}&+6\%&\b{ 0.324}&\b{0.312}&\b{0.264}&\b{\d0.279}&\b{\d0.520}&\b{0.506}&+8\%\\
\bottomrule
\end{tabular}
}
    
\end{table*}
\section{C3: Continued Pretraining with Contrastive Learning for CLIR}

In this section, we introduce C3, a continued pretraining approach with contrastive learning that encourages similar representations of a pair of texts across languages. The language model learns to establish a semantic space containing the two languages of interest with meaningful similarity by training with this objective.  

Specifically, consider a comparable corpus with linked document pairs $(d^\mathcal{S}_i, d^\mathcal{T}_i)$ in languages $\mathcal{S}$ and $\mathcal{T}$ (i.e., pairs of documents in different languages containing similar information). Given a list of such document pairs [$(d^\mathcal{S}_1, d^\mathcal{T}_1)$, $(d^\mathcal{S}_2, d^\mathcal{T}_2)$, $\dots$, $(d^\mathcal{S}_n, d^\mathcal{T}_n)$], we construct a list of spans [$s^\mathcal{S}_1$, $s^\mathcal{T}_1$, $s^\mathcal{S}_2$, $s^\mathcal{T}_2$, $\dots$, $s^\mathcal{S}_n$, $s^\mathcal{T}_n$] by randomly sampling one span from each document. 

Let $h^L_i$ be the sequence of token representations of span $s^L_i$ where $L\in \{\mathcal{S}, \mathcal{T}\}$, we construct its SimCLR~\cite{chen2020simclr} contrastive loss as 
\begin{equation*}
    \mathcal{L}^{co}_{iL} = - \log 
        \frac{ 
            \exp\left(f\left(h^\mathcal{S}_{i}, h^\mathcal{T}_{i}\right)\right)
        }{
            \sum_{j=1}^n \sum_{k\in \{\mathcal{S}, \mathcal{T}\}} 
            \mathbbm{1}(i\neq j \wedge L\neq k) 
            \exp\left(f\left(h^l_i, h^k_l\right)\right)
        }
    \label{eq:simclr}
\end{equation*}
with $\mathbbm{1}(\bullet)$ being the indicator function and $f(h_1, h_2)$ being the similarity function between representations $h_1$ and $h_2$. 
This contrastive loss is similar to the one proposed in coCondenser~\cite{gao2021unsupervised} but encourages the model to learn different knowledge. Instead of sampling pairs of spans from the same document, we construct the pair by sampling one span from each side of the linked documents. Equation~\ref{eq:simclr} promotes the representation $h^\mathcal{S}_i$ and $h^\mathcal{T}_i$ to be closer while discouraging representations of spans in the same language from being similar (since $k$ can the same as $L$).
This construction pushes the encoder away from clustering text in the same language in the semantic space and pulls the text across languages with similar meanings closer, while retaining distributional robustness by randomly matching the spans in the documents.

To promote token-level similarities, we apply the MaxSim operator proposed in ColBERT~\cite{khattab2020colbert} as the similarity function $f(h_1, h_2)$. Specifically, the function can be written as 
\begin{equation*}
    f(h_1, h_2) = \sum_{i\in |h_1|} \max_{j\in |h_2|} h_{1i}\cdot h_{2j}^T
    \label{eq:maxsim-cosine}
\end{equation*}
where $|h_\bullet|$ denotes the number of tokens in the corresponding span and $h_{\bullet k}$ denotes the representation of the $k$-th token in $h_\bullet$. With this similarity function, the contrastive learning loss flows into the token representation to explicitly promote token alignment in the semantic space. 

Finally, we combine $\mathcal{L}^{co}_{iL}$ with the masked language modeling loss $\mathcal{L}^{mlm}_{iL}$ and $\mathcal{L}^{cdmlm}_{iL}$ on span $s^L_i$ from the transformer network and the Condenser head~\cite{gao2021condenser}, respectively, to train the bottom half of the network more directly. Therefore, the total loss $\mathcal{L}$ can be expressed as 
\begin{equation*}
    \mathcal{L} = \frac{1}{2n} \sum_{i=1}^n \sum_{L\in \{\mathcal{S}, \mathcal{T}\}} 
    \left[  
        \mathcal{L}_{iL}^{co} + \mathcal{L}_{iL}^{cdmlm} + \mathcal{L}_{iL}^{mlm}
    \right]
    \label{eq:total-loss}
\end{equation*}

\textbf{\textbf{}}\section{Experiments and Analysis}

Our experiment follows the workflow in Figure~\ref{fig:pretrain-pipeline}. In this specific study, we use English as our pivot language for the queries and Chinese, Persian, French, and German as our document languages. However, we argue that C3 is generalizable to other language pairs. In the rest of the section, we discuss our experiments' data, models, setup, and results. 

\subsection{Datasets}

To continue pretraining the off-the-shelf pretrained models with C3, we assembled linked Wikipedia articles on the same topic in different languages. Specifically, we leveraged CLIRMatrix~\cite{sun2020clirmatrix}, a retrieval collection that uses the article titles as the queries to retrieve documents for 19,182 language pairs. For each language pair, we extract all query and document pairs with relevance score 6, which are the Wikipedia pages on the same topic as asserted by inter-wiki links (one query only has one document with a score of 6 given a specific language pair). These documents are linked to construct the comparable corpus. We extracted 586k, 586k, 1,283k, and 1,162k document pairs for Chinese, Persian, French, and German, respectively.

For task-specific fine-tuning, we use the ``small'' training triples provided in MSMARCO-v1, which consists of 39 million triples of query, positive, and negative passages. 

We evaluate the final retrieval models on HC4~\cite{hc4}, a newly constructed evaluation collection for CLIR, for Chinese and Persian, NTCIR~\cite{mitamura2010ntcir-aclia} for Chinese, CLEF 08-09 for Persian~\cite{clef08adhocoverview, clef09adhocoverview}, and CLEF 03 ~\cite{Braschler2003-zs} for French and German. HC4 consists of 50 topics for each language. We have 100 topics for NTCIR and CLEF 08-09 Persian and 60 topics for CLEF 03 French and German. We use the title in English as our evaluation queries.  

Despite experimenting with relatively high resource language pairs, we argue that there is no language-specific component in C3. We believe C3 is applicable to language pairs that have similar amount of linked Wikipedia pages.

\subsection{Experiment Setup}

We test our proposed approach with XLM-R-base~\cite{conneau2020unsupervised}. Additionally, we also tested XLM-align-base~\cite{chi2021xlmalign}, which is a variant of XLM-R-base pretrained with parallel text in 14 language pairs and multilingual text in 94 languages. All text in our experiments is tokenized by Sentence BPE~\cite{conneau2020unsupervised}, which XLM-R uses. 

We construct the spans from document pairs with a window of 180 tokens. We pretrain the model with C3 for 100,000 gradient update steps with an initial learning rate set to $5\times 10^{-6}$ using 4 GPUs with 24GB of memory each. We leveraged Gradient Cache~\cite{gao2021gradientcache} to run with batches of 64 document pairs (16 per GPU). 

We tested on two dense retrieval models: ColBERT~\cite{khattab2020colbert} and DPR~\cite{karpukhin2020dense}. After pretraining, each model is fine-tuned with the retrieval objective (either ColBERT or DPR) for 200,000 steps also using 4 GPUs with a learning rate set to $5\times 10^{-6}$ for each query-document language pair.  We use the original implementation of ColBERT for its fine-tuning and Tevatron~\cite{Gao2022TevatronAE} for DPR with a shared encoder for queries and documents. 
Both retrieval models are tested in a reranking setting, where all models rerank the top-1000 documents retrieved by BM25 with machine translated queries. 
The machine translation models for Chinese and Persian were trained using AWS Sockeye v2 Model~\cite{sockeye2} with 85M and 12M general domain parallel sentences for each language pair respectively. We used Google Translate for German and French.

\subsection{Results and Analysis}

\begin{table}[t]
\caption{Ablation study on different similarity function used in contrastive learning with and without the Condenser head (Cond.). The values showed in the table is nDCG@100 on HC4 Chinese test set. }\label{tab:hc4-zho-ablation}
\centering
\begin{tabular}{llc|ccc}
\toprule
        &            &       & \multicolumn{3}{c}{Contrastive Similarity}\\
Lang. Model & Ret. Model & Cond. &  None &  CLS  & MaxSim \\
\midrule
\multirow{4}{*}{XLM-R}
& \multirow{2}{*}{ColBERT }
 & \xmark &  0.352 &  0.389 &  0.410 \\
&& \cmark &     -- &  0.391 &  0.444 \\
\cmidrule{2-6}
& \multirow{2}{*}{DPR     }
 & \xmark &  0.330 &  0.382 &  0.381 \\
&& \cmark &     -- &  0.368 &  0.395 \\
\midrule
\multirow{4}{*}{XLM-A}
& \multirow{2}{*}{ColBERT }
 & \xmark &  0.425 &  0.482 &  0.474 \\
&& \cmark &     -- &  0.457 &  0.483 \\
\cmidrule{2-6}
& \multirow{2}{*}{DPR     }
 & \xmark &  0.385 &  0.406 &  0.406 \\
&& \cmark &     -- &  0.408 &  0.421 \\
\bottomrule
\end{tabular}
\end{table}

Table~\ref{tab:main-results} summarizes the main results of our experiments, which indicate that building dense retrieval models using C3 yields better effectiveness. When starting from XLM-R, C3 provides an 8\% relative improvement in nDCG@100 (and 11\% in nDCG@10) over directly fine-tuning a ColBERT model. The model benefits from the warm start before training with relevance labels by pretraining with a similar objective (MaxSim) with weakly supervised text. 
On the other hand, we observe a slightly larger gain on DPR, suggesting even retrieval models that score documents with sequence representations (i.e., embeddings of \texttt{CLS} tokens) benefit from a task that promotes token-level similarity. 

The improvement in the retrieval effectiveness by C3 is less substantial when starting from XLM-align (at most 6\% in nDCG@100 compared to 10\%). Since XLM-align is trained with parallel text, its ability to create relationships between text across languages is better than XLM-R, resulting in a diminishing return from investing computation resources in pretraining. Nevertheless, C3 still provides more effective retrieval results across languages. 

Among the evaluated language pairs, English-French is particularly interesting. Applying C3 yields negative ``improvements'' in some cases. As English and French have a close relationship linguistically, we suspect the original XLM-R model, which is not trained with parallel text, already establishes an effective cross-language semantic space.  Continued pretraining with C3 may simply not be necessary in such a case. Notably, XLM-align, which initialized its parameters by XLM-R, also yields worse retrieval results (0.590 to 0.579 in nDCG@100 and 0.514 to 0.478 in nDCG@10), which further supports our observation. 

Note that all our reranking models underperform BM25 on CLEF Persian collection. After evaluating separately on topics generated in CLEF 2008 and 2009, we discovered that the topic characteristics are different between the two (nDCG@100 of 0.421 on 08 and 0.250 on 09 for BM25). Models pretrained with C3 underperform BM25 in 2008 topics, but are at least on par with BM25 on 2009 topics. 
While this effect deserves further investigation, we note that queries for this collection were originally created in Persian and then translated into English, possibly initially by nonnative speakers~\cite{clef08adhocoverview, clef09adhocoverview}.  Perhaps the English queries in 2009 better match the English for which our models have been trained. Nevertheless, C3 still improves the pretrained language models in these cases. 

Comparing the average relative improvements (over all six test collections) that result from applying C3, we consistently see somewhat stronger relative improvements with nDCG@10 than with nDCG@100.  From this we conclude that the effects of the improved modeling are particularly helpful nearer to the top of the ranked list, where interactive users might be expected to concentrate their attention.

\begin{figure}[t]
    \centering
    \includegraphics[width=\linewidth]{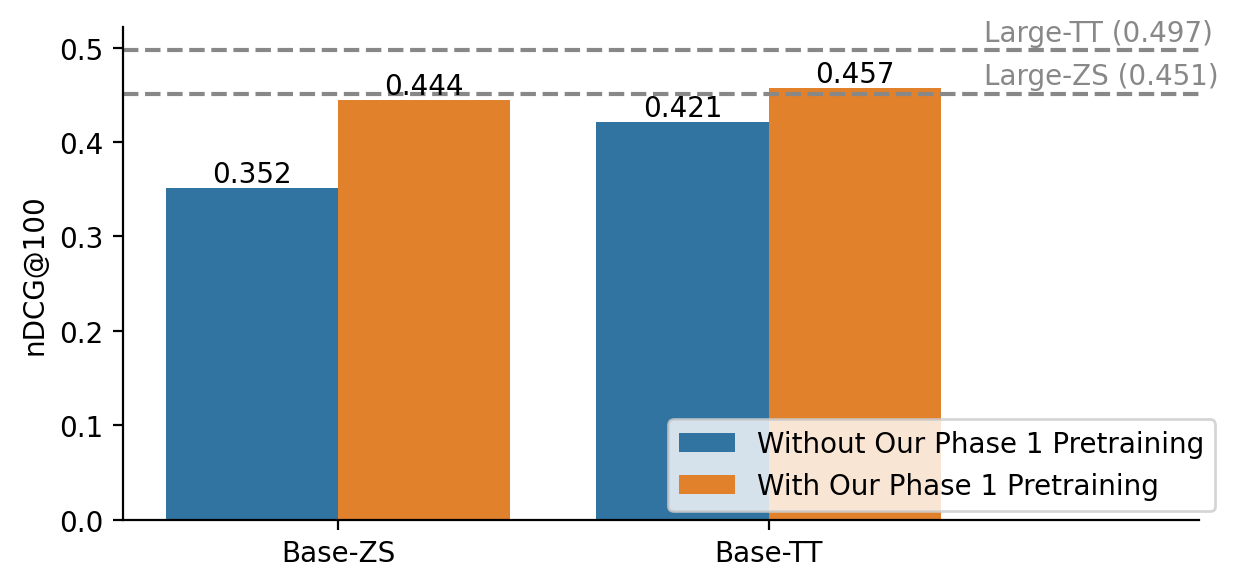}
    \caption{ColBERT models with zero-shot transfer (ZS) and translate-train (TT) approaches using XLM-RoBERTa-base on HC4 Chinese test set. The dashed line demonstrate the nDCG@100 value for XLM-RoBERTa-large with both approaches.}
    \label{fig:compare-w-large}
\end{figure}

To investigate our hypothesis regarding the utility of token-level similarity, we evaluate models in which different similarity functions were used as a basis for contrastive learning in continued pretraining. Using the CLS token in this way is similar to the coCondenser model. 
Results in Table~\ref{tab:hc4-zho-ablation} suggest that with the Condenser head, as implemented in the coCondenser model, pretraining with MaxSim similarity as the contrastive learning objective produces better retrieval models. The improvement is minimal without the Condenser head, indicating that token-level similarity benefits from routing information directly to the bottom half of the network. Interestingly, the second-best approach among the four combinations is \texttt{CLS}-based contrastive learning without using the Condenser head, which contradicts the original proposal of coCondenser. 
However, any continued pretraining is rewarding. Despite the competition among the variants, all language models with continued pretraining outperform their original off-the-shelf version. 

Finally, we ask the question: what if we can afford to translate MS MARCO so that we can use a translate-train model? 
To investigate, we utilize the Chinese translation of the MSMARCO-v1 training triples from ColBERT-X~\cite{colbertx}, which can also be accessed via \texttt{ir\_datasets}~\cite{irds} with the dataset key \texttt{neumarco/zh}\footnote{\url{https://ir-datasets.com/neumarco.html\#neumarco/zh}}.
Figure~\ref{fig:compare-w-large} shows that without C3, the ColBERT model improves from 0.352 to 0.421, which is still worse than zero-shot transfer models trained with C3 for CLIR, suggesting allocating effort to C3 rather than training a translation model when computational resources are limited. When both are affordable, the effectiveness (0.457) is on par with zero-shot transfer a ColBERT model with XLM-R-large (0.451), which is even more expensive to train. With translate-train, ColBERT with XLM-R-large achieves close to 0.5 in nDCG@100 but requires more computational resources to run.

\section{Conclusion and Future Work}

This paper proposed a continued pretraining task C3 on a weakly supervised corpus with a contrastive learning objective. We showed that the final retrieval models that are fine-tuned from models trained with C3 are more effective than off-the-shelf multilingual models. 
Further analysis suggests that translate-train can further improve retrieval models fine-tuned from C3-pretrained models. Evaluating with larger models such as XLM-R-large can also provide insight into the robustness of our approach. A natural next step would be extending C3 on some lower-resource languages where we have fewer Wikipedia articles in such languages.

Beyond that, despite being motivated by CLIR problems, C3 might also be applied to monolingual retrieval in cases where we have documents on the same topic that may use different writing styles.

\bibliographystyle{ACM-Reference-Format}
\bibliography{sample-base}


\begin{thebibliography}{44}


\ifx \showCODEN    \undefined \def \showCODEN     #1{\unskip}     \fi
\ifx \showDOI      \undefined \def \showDOI       #1{#1}\fi
\ifx \showISBNx    \undefined \def \showISBNx     #1{\unskip}     \fi
\ifx \showISBNxiii \undefined \def \showISBNxiii  #1{\unskip}     \fi
\ifx \showISSN     \undefined \def \showISSN      #1{\unskip}     \fi
\ifx \showLCCN     \undefined \def \showLCCN      #1{\unskip}     \fi
\ifx \shownote     \undefined \def \shownote      #1{#1}          \fi
\ifx \showarticletitle \undefined \def \showarticletitle #1{#1}   \fi
\ifx \showURL      \undefined \def \showURL       {\relax}        \fi
\providecommand\bibfield[2]{#2}
\providecommand\bibinfo[2]{#2}
\providecommand\natexlab[1]{#1}
\providecommand\showeprint[2][]{arXiv:#2}

\bibitem[\protect\citeauthoryear{Agirre, Nunzio, Ferro, Mandl, and
  Peters}{Agirre et~al\mbox{.}}{2008}]%
        {clef08adhocoverview}
\bibfield{author}{\bibinfo{person}{Eneko Agirre}, \bibinfo{person}{Giorgio
  Maria~Di Nunzio}, \bibinfo{person}{Nicola Ferro}, \bibinfo{person}{Thomas
  Mandl}, {and} \bibinfo{person}{Carol Peters}.}
  \bibinfo{year}{2008}\natexlab{}.
\newblock \showarticletitle{CLEF 2008: Ad hoc track overview}. In
  \bibinfo{booktitle}{\emph{Workshop of the Cross-Language Evaluation Forum for
  European Languages}}. Springer, \bibinfo{pages}{15--37}.
\newblock


\bibitem[\protect\citeauthoryear{Bajaj, Campos, Craswell, Deng, Gao, Liu,
  Majumder, McNamara, Mitra, Nguyen, Rosenberg, Song, Stoica, Tiwary, and
  Wang}{Bajaj et~al\mbox{.}}{2018}]%
        {bajaj2018ms}
\bibfield{author}{\bibinfo{person}{Payal Bajaj}, \bibinfo{person}{Daniel
  Campos}, \bibinfo{person}{Nick Craswell}, \bibinfo{person}{Li Deng},
  \bibinfo{person}{Jianfeng Gao}, \bibinfo{person}{Xiaodong Liu},
  \bibinfo{person}{Rangan Majumder}, \bibinfo{person}{Andrew McNamara},
  \bibinfo{person}{Bhaskar Mitra}, \bibinfo{person}{Tri Nguyen},
  \bibinfo{person}{Mir Rosenberg}, \bibinfo{person}{Xia Song},
  \bibinfo{person}{Alina Stoica}, \bibinfo{person}{Saurabh Tiwary}, {and}
  \bibinfo{person}{Tong Wang}.} \bibinfo{year}{2018}\natexlab{}.
\newblock \bibinfo{title}{MS MARCO: A Human Generated MAchine Reading
  COmprehension Dataset}.
\newblock
\newblock
\showeprint[arxiv]{1611.09268}~[cs.CL]


\bibitem[\protect\citeauthoryear{Bonab, Sarwar, and Allan}{Bonab
  et~al\mbox{.}}{2020}]%
        {Bonab2020-xk}
\bibfield{author}{\bibinfo{person}{Hamed Bonab},
  \bibinfo{person}{Sheikh~Muhammad Sarwar}, {and} \bibinfo{person}{James
  Allan}.} \bibinfo{year}{2020}\natexlab{}.
\newblock \showarticletitle{Training Effective Neural {CLIR} by Bridging the
  Translation Gap}.
\newblock In \bibinfo{booktitle}{\emph{Proceedings of the 43rd International
  {ACM} {SIGIR} Conference on Research and Development in Information
  Retrieval}}. \bibinfo{publisher}{Association for Computing Machinery},
  \bibinfo{address}{New York, NY, USA}, \bibinfo{pages}{9--18}.
\newblock


\bibitem[\protect\citeauthoryear{Braschler and Peters}{Braschler and
  Peters}{2004}]%
        {Braschler2003-zs}
\bibfield{author}{\bibinfo{person}{Martin Braschler} {and}
  \bibinfo{person}{Carol Peters}.} \bibinfo{year}{2004}\natexlab{}.
\newblock \showarticletitle{{CLEF} 2003 Methodology and Metrics}. In
  \bibinfo{booktitle}{\emph{Comparative Evaluation of Multilingual Information
  Access Systems}}. \bibinfo{publisher}{Springer Berlin Heidelberg},
  \bibinfo{pages}{7--20}.
\newblock


\bibitem[\protect\citeauthoryear{Chang, Yu, Chang, Yang, and Kumar}{Chang
  et~al\mbox{.}}{2020}]%
        {chang2020pre}
\bibfield{author}{\bibinfo{person}{Wei-Cheng Chang}, \bibinfo{person}{Felix~X
  Yu}, \bibinfo{person}{Yin-Wen Chang}, \bibinfo{person}{Yiming Yang}, {and}
  \bibinfo{person}{Sanjiv Kumar}.} \bibinfo{year}{2020}\natexlab{}.
\newblock \showarticletitle{Pre-training tasks for embedding-based large-scale
  retrieval}.
\newblock \bibinfo{journal}{\emph{arXiv preprint arXiv:2002.03932}}
  (\bibinfo{year}{2020}).
\newblock


\bibitem[\protect\citeauthoryear{Chen, Kornblith, Norouzi, and Hinton}{Chen
  et~al\mbox{.}}{2020}]%
        {chen2020simclr}
\bibfield{author}{\bibinfo{person}{Ting Chen}, \bibinfo{person}{Simon
  Kornblith}, \bibinfo{person}{Mohammad Norouzi}, {and}
  \bibinfo{person}{Geoffrey Hinton}.} \bibinfo{year}{2020}\natexlab{}.
\newblock \showarticletitle{A simple framework for contrastive learning of
  visual representations}. In \bibinfo{booktitle}{\emph{International
  conference on machine learning}}. PMLR, \bibinfo{pages}{1597--1607}.
\newblock


\bibitem[\protect\citeauthoryear{Chi, Dong, Zheng, Huang, Mao, Huang, and
  Wei}{Chi et~al\mbox{.}}{2021}]%
        {chi2021xlmalign}
\bibfield{author}{\bibinfo{person}{Zewen Chi}, \bibinfo{person}{Li Dong},
  \bibinfo{person}{Bo Zheng}, \bibinfo{person}{Shaohan Huang},
  \bibinfo{person}{Xian-Ling Mao}, \bibinfo{person}{Heyan Huang}, {and}
  \bibinfo{person}{Furu Wei}.} \bibinfo{year}{2021}\natexlab{}.
\newblock \showarticletitle{Improving pretrained cross-lingual language models
  via self-labeled word alignment}.
\newblock \bibinfo{journal}{\emph{arXiv preprint arXiv:2106.06381}}
  (\bibinfo{year}{2021}).
\newblock


\bibitem[\protect\citeauthoryear{Conneau, Khandelwal, Goyal, Chaudhary, Wenzek,
  Guzmán, Grave, Ott, Zettlemoyer, and Stoyanov}{Conneau
  et~al\mbox{.}}{2020}]%
        {conneau2020unsupervised}
\bibfield{author}{\bibinfo{person}{Alexis Conneau}, \bibinfo{person}{Kartikay
  Khandelwal}, \bibinfo{person}{Naman Goyal}, \bibinfo{person}{Vishrav
  Chaudhary}, \bibinfo{person}{Guillaume Wenzek}, \bibinfo{person}{Francisco
  Guzmán}, \bibinfo{person}{Edouard Grave}, \bibinfo{person}{Myle Ott},
  \bibinfo{person}{Luke Zettlemoyer}, {and} \bibinfo{person}{Veselin
  Stoyanov}.} \bibinfo{year}{2020}\natexlab{}.
\newblock \bibinfo{title}{Unsupervised Cross-lingual Representation Learning at
  Scale}.
\newblock
\newblock
\showeprint[arxiv]{1911.02116}~[cs.CL]


\bibitem[\protect\citeauthoryear{Dai and Callan}{Dai and Callan}{2019}]%
        {Dai2019-sl}
\bibfield{author}{\bibinfo{person}{Zhuyun Dai} {and} \bibinfo{person}{Jamie
  Callan}.} \bibinfo{year}{2019}\natexlab{}.
\newblock \showarticletitle{Deeper Text Understanding for {IR} with Contextual
  Neural Language Modeling}. In \bibinfo{booktitle}{\emph{Proceedings of the
  42nd International {ACM} {SIGIR} Conference on Research and Development in
  Information Retrieval}} (Paris, France) \emph{(\bibinfo{series}{SIGIR'19})}.
  \bibinfo{publisher}{Association for Computing Machinery},
  \bibinfo{address}{New York, NY, USA}, \bibinfo{pages}{985--988}.
\newblock


\bibitem[\protect\citeauthoryear{Darwish and Oard}{Darwish and Oard}{2003}]%
        {darwish2003psq}
\bibfield{author}{\bibinfo{person}{Kareem Darwish} {and}
  \bibinfo{person}{Douglas~W Oard}.} \bibinfo{year}{2003}\natexlab{}.
\newblock \showarticletitle{Probabilistic structured query methods}. In
  \bibinfo{booktitle}{\emph{Proceedings of the 26th annual international ACM
  SIGIR conference on Research and development in informaion retrieval}}.
  \bibinfo{pages}{338--344}.
\newblock


\bibitem[\protect\citeauthoryear{Devlin, Chang, Lee, and Toutanova}{Devlin
  et~al\mbox{.}}{2019}]%
        {devlin-etal-2019-bert}
\bibfield{author}{\bibinfo{person}{Jacob Devlin}, \bibinfo{person}{Ming-Wei
  Chang}, \bibinfo{person}{Kenton Lee}, {and} \bibinfo{person}{Kristina
  Toutanova}.} \bibinfo{year}{2019}\natexlab{}.
\newblock \showarticletitle{{BERT}: Pre-training of Deep Bidirectional
  Transformers for Language Understanding}. In
  \bibinfo{booktitle}{\emph{Proceedings of the 2019 Conference of the North
  {A}merican Chapter of the Association for Computational Linguistics: Human
  Language Technologies, Volume 1 (Long and Short Papers)}}.
  \bibinfo{publisher}{Association for Computational Linguistics},
  \bibinfo{address}{Minneapolis, Minnesota}, \bibinfo{pages}{4171--4186}.
\newblock
\urldef\tempurl%
\url{https://doi.org/10.18653/v1/N19-1423}
\showDOI{\tempurl}


\bibitem[\protect\citeauthoryear{Domhan, Denkowski, Vilar, Niu, Hieber, and
  Heafield}{Domhan et~al\mbox{.}}{2020}]%
        {sockeye2}
\bibfield{author}{\bibinfo{person}{Tobias Domhan}, \bibinfo{person}{Michael
  Denkowski}, \bibinfo{person}{David Vilar}, \bibinfo{person}{Xing Niu},
  \bibinfo{person}{Felix Hieber}, {and} \bibinfo{person}{Kenneth Heafield}.}
  \bibinfo{year}{2020}\natexlab{}.
\newblock \showarticletitle{The {Sockeye} 2 Neural Machine Translation Toolkit
  at {AMTA} 2020}. In \bibinfo{booktitle}{\emph{Proceedings of the 14th
  Conference of the Association for Machine Translation in the Americas (Volume
  1: Research Track)}}. \bibinfo{publisher}{Association for Machine Translation
  in the Americas}, \bibinfo{address}{Virtual}, \bibinfo{pages}{110--115}.
\newblock


\bibitem[\protect\citeauthoryear{Fan, Xie, Cai, Chen, Ma, Li, Zhang, Guo, and
  Liu}{Fan et~al\mbox{.}}{2021}]%
        {fan2021pre}
\bibfield{author}{\bibinfo{person}{Yixing Fan}, \bibinfo{person}{Xiaohui Xie},
  \bibinfo{person}{Yinqiong Cai}, \bibinfo{person}{Jia Chen},
  \bibinfo{person}{Xinyu Ma}, \bibinfo{person}{Xiangsheng Li},
  \bibinfo{person}{Ruqing Zhang}, \bibinfo{person}{Jiafeng Guo}, {and}
  \bibinfo{person}{Yiqun Liu}.} \bibinfo{year}{2021}\natexlab{}.
\newblock \showarticletitle{Pre-training Methods in Information Retrieval}.
\newblock \bibinfo{journal}{\emph{arXiv preprint arXiv:2111.13853}}
  (\bibinfo{year}{2021}).
\newblock


\bibitem[\protect\citeauthoryear{Ferro and Peters}{Ferro and Peters}{2009}]%
        {clef09adhocoverview}
\bibfield{author}{\bibinfo{person}{Nicola Ferro} {and} \bibinfo{person}{Carol
  Peters}.} \bibinfo{year}{2009}\natexlab{}.
\newblock \showarticletitle{CLEF 2009 ad hoc track overview: TEL and persian
  tasks}. In \bibinfo{booktitle}{\emph{Workshop of the Cross-Language
  Evaluation Forum for European Languages}}. Springer, \bibinfo{pages}{13--35}.
\newblock


\bibitem[\protect\citeauthoryear{Gao and Callan}{Gao and Callan}{2021a}]%
        {gao2021condenser}
\bibfield{author}{\bibinfo{person}{Luyu Gao} {and} \bibinfo{person}{Jamie
  Callan}.} \bibinfo{year}{2021}\natexlab{a}.
\newblock \showarticletitle{Condenser: a Pre-training Architecture for Dense
  Retrieval}.
\newblock \bibinfo{journal}{\emph{arXiv preprint arXiv:2104.08253}}
  (\bibinfo{year}{2021}).
\newblock


\bibitem[\protect\citeauthoryear{Gao and Callan}{Gao and Callan}{2021b}]%
        {gao2021unsupervised}
\bibfield{author}{\bibinfo{person}{Luyu Gao} {and} \bibinfo{person}{Jamie
  Callan}.} \bibinfo{year}{2021}\natexlab{b}.
\newblock \showarticletitle{Unsupervised corpus aware language model
  pre-training for dense passage retrieval}.
\newblock \bibinfo{journal}{\emph{arXiv preprint arXiv:2108.05540}}
  (\bibinfo{year}{2021}).
\newblock


\bibitem[\protect\citeauthoryear{Gao, Ma, Lin, and Callan}{Gao
  et~al\mbox{.}}{2022}]%
        {Gao2022TevatronAE}
\bibfield{author}{\bibinfo{person}{Luyu Gao}, \bibinfo{person}{Xueguang Ma},
  \bibinfo{person}{Jimmy~J. Lin}, {and} \bibinfo{person}{Jamie Callan}.}
  \bibinfo{year}{2022}\natexlab{}.
\newblock \showarticletitle{Tevatron: An Efficient and Flexible Toolkit for
  Dense Retrieval}.
\newblock \bibinfo{journal}{\emph{ArXiv}}  \bibinfo{volume}{abs/2203.05765}
  (\bibinfo{year}{2022}).
\newblock


\bibitem[\protect\citeauthoryear{Gao, Zhang, Han, and Callan}{Gao
  et~al\mbox{.}}{2021}]%
        {gao2021gradientcache}
\bibfield{author}{\bibinfo{person}{Luyu Gao}, \bibinfo{person}{Yunyi Zhang},
  \bibinfo{person}{Jiawei Han}, {and} \bibinfo{person}{Jamie Callan}.}
  \bibinfo{year}{2021}\natexlab{}.
\newblock \showarticletitle{Scaling deep contrastive learning batch size under
  memory limited setup}.
\newblock \bibinfo{journal}{\emph{arXiv preprint arXiv:2101.06983}}
  (\bibinfo{year}{2021}).
\newblock


\bibitem[\protect\citeauthoryear{Huang, Bonab, Sarwar, Rahimi, and Allan}{Huang
  et~al\mbox{.}}{2021}]%
        {huang2021mixed}
\bibfield{author}{\bibinfo{person}{Zhiqi Huang}, \bibinfo{person}{Hamed Bonab},
  \bibinfo{person}{Sheikh~Muhammad Sarwar}, \bibinfo{person}{Razieh Rahimi},
  {and} \bibinfo{person}{James Allan}.} \bibinfo{year}{2021}\natexlab{}.
\newblock \bibinfo{booktitle}{\emph{Mixed Attention Transformer for Leveraging
  Word-Level Knowledge to Neural Cross-Lingual Information Retrieval}}.
\newblock \bibinfo{publisher}{Association for Computing Machinery},
  \bibinfo{address}{New York, NY, USA}, \bibinfo{pages}{760–770}.
\newblock
\showISBNx{9781450384469}
\urldef\tempurl%
\url{https://doi.org/10.1145/3459637.3482452}
\showURL{%
\tempurl}


\bibitem[\protect\citeauthoryear{Izacard, Caron, Hosseini, Riedel, Bojanowski,
  Joulin, and Grave}{Izacard et~al\mbox{.}}{2021a}]%
        {izacard2021unsupervised}
\bibfield{author}{\bibinfo{person}{Gautier Izacard}, \bibinfo{person}{Mathilde
  Caron}, \bibinfo{person}{Lucas Hosseini}, \bibinfo{person}{Sebastian Riedel},
  \bibinfo{person}{Piotr Bojanowski}, \bibinfo{person}{Armand Joulin}, {and}
  \bibinfo{person}{Edouard Grave}.} \bibinfo{year}{2021}\natexlab{a}.
\newblock \bibinfo{title}{Towards Unsupervised Dense Information Retrieval with
  Contrastive Learning}.
\newblock
\newblock
\showeprint[arxiv]{2112.09118}~[cs.IR]


\bibitem[\protect\citeauthoryear{Izacard, Caron, Hosseini, Riedel, Bojanowski,
  Joulin, and Grave}{Izacard et~al\mbox{.}}{2021b}]%
        {izacard2021towards}
\bibfield{author}{\bibinfo{person}{Gautier Izacard}, \bibinfo{person}{Mathilde
  Caron}, \bibinfo{person}{Lucas Hosseini}, \bibinfo{person}{Sebastian Riedel},
  \bibinfo{person}{Piotr Bojanowski}, \bibinfo{person}{Armand Joulin}, {and}
  \bibinfo{person}{Edouard Grave}.} \bibinfo{year}{2021}\natexlab{b}.
\newblock \showarticletitle{Towards Unsupervised Dense Information Retrieval
  with Contrastive Learning}.
\newblock \bibinfo{journal}{\emph{arXiv preprint arXiv:2112.09118}}
  (\bibinfo{year}{2021}).
\newblock


\bibitem[\protect\citeauthoryear{Jiang, El-Jaroudi, Hartmann, Karakos, and
  Zhao}{Jiang et~al\mbox{.}}{2020}]%
        {Jiang2020-dt}
\bibfield{author}{\bibinfo{person}{Zhuolin Jiang}, \bibinfo{person}{Amro
  El-Jaroudi}, \bibinfo{person}{William Hartmann}, \bibinfo{person}{Damianos
  Karakos}, {and} \bibinfo{person}{Lingjun Zhao}.}
  \bibinfo{year}{2020}\natexlab{}.
\newblock \showarticletitle{Cross-lingual Information Retrieval with {BERT}}.
\newblock  (\bibinfo{date}{April} \bibinfo{year}{2020}).
\newblock
\showeprint[arxiv]{2004.13005}~[cs.IR]


\bibitem[\protect\citeauthoryear{Karpukhin, Oğuz, Min, Lewis, Wu, Edunov,
  Chen, and tau Yih}{Karpukhin et~al\mbox{.}}{2020}]%
        {karpukhin2020dense}
\bibfield{author}{\bibinfo{person}{Vladimir Karpukhin}, \bibinfo{person}{Barlas
  Oğuz}, \bibinfo{person}{Sewon Min}, \bibinfo{person}{Patrick Lewis},
  \bibinfo{person}{Ledell Wu}, \bibinfo{person}{Sergey Edunov},
  \bibinfo{person}{Danqi Chen}, {and} \bibinfo{person}{Wen tau Yih}.}
  \bibinfo{year}{2020}\natexlab{}.
\newblock \bibinfo{title}{Dense Passage Retrieval for Open-Domain Question
  Answering}.
\newblock
\newblock
\showeprint[arxiv]{2004.04906}~[cs.CL]


\bibitem[\protect\citeauthoryear{Khattab and Zaharia}{Khattab and
  Zaharia}{2020}]%
        {khattab2020colbert}
\bibfield{author}{\bibinfo{person}{Omar Khattab} {and} \bibinfo{person}{Matei
  Zaharia}.} \bibinfo{year}{2020}\natexlab{}.
\newblock \bibinfo{title}{ColBERT: Efficient and Effective Passage Search via
  Contextualized Late Interaction over BERT}.
\newblock
\newblock
\showeprint[arxiv]{2004.12832}~[cs.IR]


\bibitem[\protect\citeauthoryear{Kwiatkowski, Palomaki, Redfield, Collins,
  Parikh, Alberti, Epstein, Polosukhin, Devlin, Lee, Toutanova, Jones, Kelcey,
  Chang, Dai, Uszkoreit, Le, and Petrov}{Kwiatkowski et~al\mbox{.}}{2019}]%
        {kwiatkowski-etal-2019-natural}
\bibfield{author}{\bibinfo{person}{Tom Kwiatkowski},
  \bibinfo{person}{Jennimaria Palomaki}, \bibinfo{person}{Olivia Redfield},
  \bibinfo{person}{Michael Collins}, \bibinfo{person}{Ankur Parikh},
  \bibinfo{person}{Chris Alberti}, \bibinfo{person}{Danielle Epstein},
  \bibinfo{person}{Illia Polosukhin}, \bibinfo{person}{Jacob Devlin},
  \bibinfo{person}{Kenton Lee}, \bibinfo{person}{Kristina Toutanova},
  \bibinfo{person}{Llion Jones}, \bibinfo{person}{Matthew Kelcey},
  \bibinfo{person}{Ming-Wei Chang}, \bibinfo{person}{Andrew~M. Dai},
  \bibinfo{person}{Jakob Uszkoreit}, \bibinfo{person}{Quoc Le}, {and}
  \bibinfo{person}{Slav Petrov}.} \bibinfo{year}{2019}\natexlab{}.
\newblock \showarticletitle{Natural Questions: A Benchmark for Question
  Answering Research}.
\newblock \bibinfo{journal}{\emph{Transactions of the Association for
  Computational Linguistics}}  \bibinfo{volume}{7} (\bibinfo{year}{2019}),
  \bibinfo{pages}{452--466}.
\newblock
\urldef\tempurl%
\url{https://doi.org/10.1162/tacl_a_00276}
\showDOI{\tempurl}


\bibitem[\protect\citeauthoryear{Lawrie, Mayfield, Oard, and Yang}{Lawrie
  et~al\mbox{.}}{2022}]%
        {hc4}
\bibfield{author}{\bibinfo{person}{Dawn Lawrie}, \bibinfo{person}{James
  Mayfield}, \bibinfo{person}{Douglas~W. Oard}, {and} \bibinfo{person}{Eugene
  Yang}.} \bibinfo{year}{2022}\natexlab{}.
\newblock \showarticletitle{HC4: A New Suite of Test Collections for Ad Hoc
  CLIR}.
\newblock  (\bibinfo{year}{2022}).
\newblock
\urldef\tempurl%
\url{https://arxiv.org/abs/2201.09992}
\showURL{%
\tempurl}


\bibitem[\protect\citeauthoryear{Litschko, Vuli{\'c}, Ponzetto, and
  Glava{\v{s}}}{Litschko et~al\mbox{.}}{2021}]%
        {litschko2021cross}
\bibfield{author}{\bibinfo{person}{Robert Litschko}, \bibinfo{person}{Ivan
  Vuli{\'c}}, \bibinfo{person}{Simone~Paolo Ponzetto}, {and}
  \bibinfo{person}{Goran Glava{\v{s}}}.} \bibinfo{year}{2021}\natexlab{}.
\newblock \showarticletitle{On Cross-Lingual Retrieval with Multilingual Text
  Encoders}.
\newblock \bibinfo{journal}{\emph{arXiv preprint arXiv:2112.11031}}
  (\bibinfo{year}{2021}).
\newblock


\bibitem[\protect\citeauthoryear{MacAvaney, Soldaini, and Goharian}{MacAvaney
  et~al\mbox{.}}{2020}]%
        {macavaney2020newdog}
\bibfield{author}{\bibinfo{person}{Sean MacAvaney}, \bibinfo{person}{Luca
  Soldaini}, {and} \bibinfo{person}{Nazli Goharian}.}
  \bibinfo{year}{2020}\natexlab{}.
\newblock \showarticletitle{Teaching a New Dog Old Tricks: Resurrecting
  Multilingual Retrieval Using Zero-shot Learning}. In
  \bibinfo{booktitle}{\emph{Proceedings of the 42nd European Conference on
  Information Retrieval Research}}. \bibinfo{pages}{246--254}.
\newblock
\urldef\tempurl%
\url{https://doi.org/10.1007/978-3-030-45442-5_31}
\showDOI{\tempurl}


\bibitem[\protect\citeauthoryear{MacAvaney, Yates, Cohan, and
  Goharian}{MacAvaney et~al\mbox{.}}{2019}]%
        {macavaney2019cedr}
\bibfield{author}{\bibinfo{person}{Sean MacAvaney}, \bibinfo{person}{Andrew
  Yates}, \bibinfo{person}{Arman Cohan}, {and} \bibinfo{person}{Nazli
  Goharian}.} \bibinfo{year}{2019}\natexlab{}.
\newblock \showarticletitle{CEDR: Contextualized embeddings for document
  ranking}. In \bibinfo{booktitle}{\emph{Proceedings of the 42nd International
  ACM SIGIR Conference on Research and Development in Information Retrieval}}.
  \bibinfo{pages}{1101--1104}.
\newblock


\bibitem[\protect\citeauthoryear{MacAvaney, Yates, Feldman, Downey, Cohan, and
  Goharian}{MacAvaney et~al\mbox{.}}{2021}]%
        {irds}
\bibfield{author}{\bibinfo{person}{Sean MacAvaney}, \bibinfo{person}{Andrew
  Yates}, \bibinfo{person}{Sergey Feldman}, \bibinfo{person}{Doug Downey},
  \bibinfo{person}{Arman Cohan}, {and} \bibinfo{person}{Nazli Goharian}.}
  \bibinfo{year}{2021}\natexlab{}.
\newblock \showarticletitle{Simplified Data Wrangling with ir\_datasets}. In
  \bibinfo{booktitle}{\emph{Proceedings of the 44th International ACM SIGIR
  Conference on Research and Development in Information Retrieval}}.
  \bibinfo{pages}{2429--2436}.
\newblock


\bibitem[\protect\citeauthoryear{MITAMURA}{MITAMURA}{2010}]%
        {mitamura2010ntcir-aclia}
\bibfield{author}{\bibinfo{person}{T MITAMURA}.}
  \bibinfo{year}{2010}\natexlab{}.
\newblock \showarticletitle{Overview of the NTCIR-8 ACLIA Tasks: Advanced
  cross-lingual information access}. In \bibinfo{booktitle}{\emph{NTCIR-8
  Workshop, 2010}}.
\newblock


\bibitem[\protect\citeauthoryear{Nair, Yang, Lawrie, Duh, McNamee, Murray,
  Mayfield, and Oard}{Nair et~al\mbox{.}}{2022}]%
        {colbertx}
\bibfield{author}{\bibinfo{person}{Suraj Nair}, \bibinfo{person}{Eugene Yang},
  \bibinfo{person}{Dawn Lawrie}, \bibinfo{person}{Kevin Duh},
  \bibinfo{person}{Paul McNamee}, \bibinfo{person}{Kenton Murray},
  \bibinfo{person}{James Mayfield}, {and} \bibinfo{person}{Douglas~W. Oard}.}
  \bibinfo{year}{2022}\natexlab{}.
\newblock \showarticletitle{Transfer Learning Approaches for Building
  Cross-Language Dense Retrieval Models}.
\newblock  (\bibinfo{year}{2022}).
\newblock
\urldef\tempurl%
\url{https://arxiv.org/abs/2201.08471}
\showURL{%
\tempurl}


\bibitem[\protect\citeauthoryear{Nogueira, Yang, Cho, and Lin}{Nogueira
  et~al\mbox{.}}{2019}]%
        {nogueira2019multistage}
\bibfield{author}{\bibinfo{person}{Rodrigo Nogueira}, \bibinfo{person}{Wei
  Yang}, \bibinfo{person}{Kyunghyun Cho}, {and} \bibinfo{person}{Jimmy Lin}.}
  \bibinfo{year}{2019}\natexlab{}.
\newblock \bibinfo{title}{Multi-Stage Document Ranking with BERT}.
\newblock
\newblock
\showeprint[arxiv]{1910.14424}~[cs.IR]


\bibitem[\protect\citeauthoryear{Oord, Li, and Vinyals}{Oord
  et~al\mbox{.}}{2018}]%
        {oord2018infonce}
\bibfield{author}{\bibinfo{person}{Aaron van~den Oord}, \bibinfo{person}{Yazhe
  Li}, {and} \bibinfo{person}{Oriol Vinyals}.} \bibinfo{year}{2018}\natexlab{}.
\newblock \showarticletitle{Representation learning with contrastive predictive
  coding}.
\newblock \bibinfo{journal}{\emph{arXiv preprint arXiv:1807.03748}}
  (\bibinfo{year}{2018}).
\newblock


\bibitem[\protect\citeauthoryear{Qu, Ding, Liu, Liu, Ren, Zhao, Dong, Wu, and
  Wang}{Qu et~al\mbox{.}}{2020}]%
        {qu2020rocketqa}
\bibfield{author}{\bibinfo{person}{Yingqi Qu}, \bibinfo{person}{Yuchen Ding},
  \bibinfo{person}{Jing Liu}, \bibinfo{person}{Kai Liu},
  \bibinfo{person}{Ruiyang Ren}, \bibinfo{person}{Wayne~Xin Zhao},
  \bibinfo{person}{Daxiang Dong}, \bibinfo{person}{Hua Wu}, {and}
  \bibinfo{person}{Haifeng Wang}.} \bibinfo{year}{2020}\natexlab{}.
\newblock \showarticletitle{RocketQA: An optimized training approach to dense
  passage retrieval for open-domain question answering}.
\newblock \bibinfo{journal}{\emph{arXiv preprint arXiv:2010.08191}}
  (\bibinfo{year}{2020}).
\newblock


\bibitem[\protect\citeauthoryear{Robertson, Walker, Jones, Hancock-Beaulieu,
  and Gatford}{Robertson et~al\mbox{.}}{1994}]%
        {Robertson1994OkapiAT}
\bibfield{author}{\bibinfo{person}{Stephen~E. Robertson},
  \bibinfo{person}{Steve Walker}, \bibinfo{person}{Susan Jones},
  \bibinfo{person}{Micheline Hancock-Beaulieu}, {and} \bibinfo{person}{Mike
  Gatford}.} \bibinfo{year}{1994}\natexlab{}.
\newblock \showarticletitle{Okapi at TREC-3}. In
  \bibinfo{booktitle}{\emph{TREC}}.
\newblock


\bibitem[\protect\citeauthoryear{Shi and Lin}{Shi and Lin}{2019a}]%
        {shi2019cross}
\bibfield{author}{\bibinfo{person}{Peng Shi} {and} \bibinfo{person}{Jimmy
  Lin}.} \bibinfo{year}{2019}\natexlab{a}.
\newblock \showarticletitle{Cross-lingual relevance transfer for document
  retrieval}.
\newblock \bibinfo{journal}{\emph{arXiv preprint arXiv:1911.02989}}
  (\bibinfo{year}{2019}).
\newblock


\bibitem[\protect\citeauthoryear{Shi and Lin}{Shi and Lin}{2019b}]%
        {Shi2019-on}
\bibfield{author}{\bibinfo{person}{P Shi} {and} \bibinfo{person}{J Lin}.}
  \bibinfo{year}{2019}\natexlab{b}.
\newblock \showarticletitle{Cross-lingual relevance transfer for document
  retrieval}.
\newblock \bibinfo{journal}{\emph{arXiv preprint arXiv:1911.02989}}
  (\bibinfo{year}{2019}).
\newblock


\bibitem[\protect\citeauthoryear{Shi, Zhang, Bai, and Lin}{Shi
  et~al\mbox{.}}{2021}]%
        {shi2021cross}
\bibfield{author}{\bibinfo{person}{Peng Shi}, \bibinfo{person}{Rui Zhang},
  \bibinfo{person}{He Bai}, {and} \bibinfo{person}{Jimmy Lin}.}
  \bibinfo{year}{2021}\natexlab{}.
\newblock \showarticletitle{Cross-lingual training with dense retrieval for
  document retrieval}.
\newblock \bibinfo{journal}{\emph{arXiv preprint arXiv:2109.01628}}
  (\bibinfo{year}{2021}).
\newblock


\bibitem[\protect\citeauthoryear{Sun and Duh}{Sun and Duh}{2020}]%
        {sun2020clirmatrix}
\bibfield{author}{\bibinfo{person}{Shuo Sun} {and} \bibinfo{person}{Kevin
  Duh}.} \bibinfo{year}{2020}\natexlab{}.
\newblock \showarticletitle{CLIRMatrix: A massively large collection of
  bilingual and multilingual datasets for Cross-Lingual Information Retrieval}.
  In \bibinfo{booktitle}{\emph{Proceedings of the 2020 Conference on Empirical
  Methods in Natural Language Processing (EMNLP)}}.
  \bibinfo{pages}{4160--4170}.
\newblock


\bibitem[\protect\citeauthoryear{Xiong, Xiong, Li, Tang, Liu, Bennett, Ahmed,
  and Overwijk}{Xiong et~al\mbox{.}}{2020}]%
        {xiong2020approximate}
\bibfield{author}{\bibinfo{person}{Lee Xiong}, \bibinfo{person}{Chenyan Xiong},
  \bibinfo{person}{Ye Li}, \bibinfo{person}{Kwok-Fung Tang},
  \bibinfo{person}{Jialin Liu}, \bibinfo{person}{Paul Bennett},
  \bibinfo{person}{Junaid Ahmed}, {and} \bibinfo{person}{Arnold Overwijk}.}
  \bibinfo{year}{2020}\natexlab{}.
\newblock \bibinfo{title}{Approximate Nearest Neighbor Negative Contrastive
  Learning for Dense Text Retrieval}.
\newblock
\newblock
\showeprint[arxiv]{2007.00808}~[cs.IR]


\bibitem[\protect\citeauthoryear{Yu and Allan}{Yu and Allan}{2020}]%
        {Yu2020-dl}
\bibfield{author}{\bibinfo{person}{P Yu} {and} \bibinfo{person}{J Allan}.}
  \bibinfo{year}{2020}\natexlab{}.
\newblock \showarticletitle{A Study of Neural Matching Models for Cross-lingual
  {IR}}.
\newblock \bibinfo{journal}{\emph{Proceedings of the 43rd International ACM
  SIGIR}} (\bibinfo{year}{2020}).
\newblock


\bibitem[\protect\citeauthoryear{Yu, Fei, and Li}{Yu et~al\mbox{.}}{2021}]%
        {yu2021cross}
\bibfield{author}{\bibinfo{person}{Puxuan Yu}, \bibinfo{person}{Hongliang Fei},
  {and} \bibinfo{person}{Ping Li}.} \bibinfo{year}{2021}\natexlab{}.
\newblock \showarticletitle{Cross-lingual Language Model Pretraining for
  Retrieval}. In \bibinfo{booktitle}{\emph{Proceedings of the Web Conference
  2021}}. \bibinfo{pages}{1029--1039}.
\newblock


\bibitem[\protect\citeauthoryear{Zhao, Zbib, Jiang, Karakos, and Huang}{Zhao
  et~al\mbox{.}}{2019}]%
        {zhao2019weakly}
\bibfield{author}{\bibinfo{person}{Lingjun Zhao}, \bibinfo{person}{Rabih Zbib},
  \bibinfo{person}{Zhuolin Jiang}, \bibinfo{person}{Damianos Karakos}, {and}
  \bibinfo{person}{Zhongqiang Huang}.} \bibinfo{year}{2019}\natexlab{}.
\newblock \showarticletitle{Weakly supervised attentional model for low
  resource ad-hoc cross-lingual information retrieval}. In
  \bibinfo{booktitle}{\emph{Proceedings of the 2nd Workshop on Deep Learning
  Approaches for Low-Resource NLP (DeepLo 2019)}}. \bibinfo{pages}{259--264}.
\newblock


\end{thebibliography}

\end{document}